\documentclass[aps,prl,reprint,groupedaddress, nofootinbib]{revtex4-1}
\usepackage{epsfig}
\usepackage{graphicx}
\usepackage{color}
\begin{document}
%\vspace {1.0 in}

\title{PMN: a minimal induced-moment soft pseudo-spin glass perspective}

\author{David Sherrington}
\email[]{D.Sherrington1@physics.ox.ac.uk}

\affiliation{ Rudolf Peierls Centre for Theoretical Physics, University of Oxford, 
1 Keble Road, Oxford OX1 3NP, UK\\
Santa Fe Institute, 1399 Hyde Park Rd., Santa Fe, NM 87501, USA} 

\date{30 January 2014}

%\maketitle

\begin{abstract}
An argument that  relaxor ferroelectricity in the isovalent alloy $\mathrm {Ba(Zr}_{1-x}\mathrm{Ti}_{x})\mathrm{O}_3$ can be understood as an induced moment soft pseudo-spin glass on the $B$-ions of the $AB\mathrm{O}_{3}$  matrix \cite{DS13} is extended to the experimentally paradigmic but theoretically more complex heterovalent  relaxor $\mathrm {Pb(Mg}_{1/3}\mathrm{Nb}_{2/3}\mathrm{)O}_3$ (PMN).
It is argued that interesting  behaviour of the onset of non-ergodicity, frequency-dependent permittivity peaks and precursor nanodomains  can be understood from analagous considerations of the $B$-ions, with the displacements of the Pb ions a largely independent, but distracting, side-feature. This contrasts with conventional conceptualizations.
\end{abstract}
\maketitle

For many years, 
following 
the discovery 
over 50 years ago \cite{Smolenskii2} 
of frequency-dependent peaking as a function of temperature of the permittivity of the $AB\mathrm{O}_{3}$ 
perovskite alloy $\mathrm {Pb(Mg}_{1/3}\mathrm{Nb}_{2/3}\mathrm{)O}_3$ (PMN) , there has been much interest in relaxor ferroelectricity \cite{Cross}; for recent reviews see \cite{Cowley, Bokov, Samara}. 
However, a fully accepted microscopic understanding remains elusive. 

In a recent brief paper \cite{DS13} it was argued that several features of the 
isovalent \cite{foot-isovalent}
relaxor alloy $\mathrm {Ba}(\mathrm{Zr}_{1-x}\mathrm{Ti}_{x}\mathrm{)O}_3$ (BZT), found in extensive first principles simulations \cite{Akbarzadeh2012} and in experiments \cite{Maiti, Shvartsman}, can be understood qualitatively in terms of a simple mapping to an effective induced-moment soft pseudo-spin glass. In this paper this picture is extended to consider the more famous but also more complex 
non-isovalent
PMN system.
The intent is to try to pick out key features driving the relaxor behaviour and the observation of nanodomains, to provide possible explanations of these and other observations and to identify similarities and differences compared to BZT.

The generic pure $AB\mathrm{O}_{3}$ perovskite ferroelectrics are ionic crystals in which the $A$ ions ({\it{e.g.}} Ba, Sr, Pb) have nominal charge +2, the $B$-ions ({\it{e.g.}} Ti) have charge +4  and the O have charge -2. 
Zr in $\mathrm {Ba}\mathrm{Zr}\mathrm{O}_3$ 
also has charge +4, although it does not exhibit ferroelecticity, so there is no significant  change of charge distribution on alloying $\mathrm {Ba}\mathrm{Ti}\mathrm{O}_3$ 
 to  $\mathrm {Ba}(\mathrm{Zr}_{1-x}\mathrm{Ti}_{x}\mathrm{)O}_3$. Ti in $\mathrm {Pb}\mathrm{Ti}\mathrm{O}_3$ (PT) also has charge +4, but Mg and Nb, which randomly replace it in PMN, have respectively charges +2 and +5,  perturbations relative to PT respectively of -2 and +1.  This paper is concerned with the consequential modifications to the behavior to be expected from an extension of the mapping in \cite{DS13}. It is argued below that the origin of the relaxor behavior and the observed non-ergodicity in PMN is  qualitatively as for BZT,  
 driven by $B$-site distortions, but with also (i) additional inter-$B$-site quasi-exchange, (ii)  random field-like forces on $B$-sites \cite{foot-B} and (iii) changes in the local Pb and O site distributions that are not qualitatively relevant for the relaxor behaviour.

The methodology to be employed is to model the system with a Hamiltonian consisting of two parts, the first an analogue of that of BZT such as would correspond to occupation of the $B$-sites by fictitious ions Mg* and Nb* similar  to Mg and Nb but with charges +4, and the second  representing the difference due to the real Mg and Nb with charges +2 and +5.

For orientational purposes it may be helpful to recall the argument for BZT. In the generic  $AB\mathrm{O}_{3}$ nomenclature it was argued that for the qualitative understanding of relaxor behaviour  (i) one can usefully consider modeling in terms of  ionic displacements only on the $B$-sites, absorbing the effects of the $A$ and O ions into an effective $B$-$B$ interaction term that is spatially frustrated as a function of site-separation \cite{Zhong} but dominated by a ferroelectric compromise for pure $\mathrm {Ba}\mathrm{Ti}\mathrm{O}_3$, and (ii)  the most important aspect of the alloying lies in a  difference in the strengths of  local harmonic restoring forces at  $B$-sites depending upon whether they are occupied by  Ti or  Zr ions \cite{foot-caveats}. 
Ignoring local anisotropy, for conceptual simplicity, one is then left with a model characterized by an effective $B$-site Hamiltonian 
\begin{eqnarray}
H^{B}_{BZT}= & \sum_{i(B)} & \{{\kappa_{i}{|\bf{u_{i}}|^2}} + {\lambda_{i}}{|\bf{u_{i}}}]^4\}  \nonumber
\\  & + &  {\sum_{ij} H^{avg}_{int}({\bf{u}}_{i}, {\bf{u}}_{j},  {\bf{R}}_{ij})}
\label{eq:H2}
\end{eqnarray}
where the $\{{\bf{u}_{i}}\}$ are the local displacements, the sites $\{i\}$  are occupied randomly by Ti or Zr (according to the admixture concentration)  with corresponding  $\kappa$ and $\lambda$,  the first two terms represent the local restoring energies  and the $H_{int}$ term includes all other relevant ($B$-$B$) interaction terms. Ordered states, different from paraelectric, arise when bootstrapping of finite $\{{\bf{u}}_{i}\}$ leads to sufficient negative interaction energy to overcome the (positive energy) penalty arising from the local distortion term; the actual ordered state is that with the lowest free energy. As noted above,
the interaction part of the Hamiltonian is frustrated (competitive in ordering preferences) as a function of site-separation, with both ferroelectric and antiferroelectric contributions \cite{Zhong,DS13}. 

For pure  $\mathrm {Ba}\mathrm{Ti}\mathrm{O}_3$ the local Ti restoring force term is weak enough for such cooperative ordering to occur, whereas in  $\mathrm {Ba}\mathrm{Zr}\mathrm{O}_3$ the corresponding Zr restoring term is too strong, permitting only   paraelectricity. As demonstrated by experiment, with every $B$-site occupied by Ti the best compromise/lowest energy ordered state is ferroelectric, despite the presence also of antiferroelectric interaction terms in $H_{int}$.
On the other hand, for random $B$-site occupation by two different types of ion the cooperative ordered state need not be ferroelectric and indeed for $x<x_c \sim 0.7$ \cite{Kleemann-Miga}  in  $\mathrm{BaZr}_{1-x}\mathrm{Ti}_{x}\mathrm{O}_3$ relaxor behaviour ensues, argued \cite{DS13} as due to the ordering preference being  pseudo-spin glass, driven by bootstrapping of (only) Ti ion displacements, in an induced-moment soft-pseudospin extended analogy with conventional spin glass alloys such as $\rm{ Au}_{1-x}\rm{Fe}_{x} $ \cite{Mydosh} {\cite{foot-sg}}.   

 A similar reasoning can be applied to the model (fictitious) random alloy 
$\mathrm {Pb}(\mathrm{Mg^{*}}_{x}\mathrm{Nb^{*}}_{1-x}\mathrm{)O}_3$ (PM*N*)
in which Mg* and Nb* are  fictitious ions having the same charges as Ti (+4) but the 
local distortion coefficients of Mg and Nb, randomly distributed on the $B$-sites according to the weighting indicated. The same argument as for BZT \cite{DS13} leads to the effective $B$-site Hamiltonian 
\begin{eqnarray}
H^{B}_{PM^{*}N^{*}}= & \sum_{i} &  \{{\kappa_{i}{|\bf{u_{i}}|^2}} + {\lambda_{i}}{|\bf{u_{i}}}]^4\}   \nonumber \\  & + &
{\sum_{ij} H^{PM^{*}N^{*}}_{int}({\bf{u}}_{i}, {\bf{u}}_{j},  {\bf{R}}_{ij})}.
\label{eq:H3}
\end{eqnarray}
Both from  comparison of ionic radii \cite{foot-radii} and from  first-principles calculations \cite{Bellaiche-Vanderbilt2} it is expected 
that $\kappa^{Nb}$ will be small enough for Nb* to be relevant for cooperative ordering while $\kappa^{Mg}$ will be too large for bootstrapping displacements of Mg* ions, which will consequently be effectively frozen, respectively  analagously to the situations of Ti and Zr in BZT. Thus the relevant part of this Hamiltonian can be re-written as
\begin{eqnarray}
H^{B}_{PM^{*}N^{*}}= & \sum_{i, {\rm{Nb}}} &  \{{\kappa^{Nb^{*}}{|\bf{u_{i}}|^2}} + {\lambda^{\bf{Nb^*}}}{|\bf{u_{i}}}]^4\}   \nonumber \\  & + &
{\sum_{ij,{\rm{Nb}}} H^{PM^{*}N^{*}}_{int}({\bf{u}}_{i}, {\bf{u}}_{j},  {\bf{R}}_{ij})}
\label{eq:H4}
\end{eqnarray}
with the  summations restricted to Nb sites. Consequently, we would expect PM*N* to have properties very similar to those of BZT. While PM*N* can, in principle, be considered for a range of $x$, it is interesting to note that $x=2/3$, as in PMN, is close to the top end of the range of $x$-values giving relaxor behavior in BZT, at which the proximity of the ferroelectic phase results in significant enhancement of the permittivity peak. By comparison, PMN also has a high permittivity peak.

Since the interaction term is expected to have a similarly competitive character to that in $\mathrm{BaTiO}_{3}$ (or BZT), one might reasonably expect this fictitious alloy to show an ordering transition from ferroelectric to relaxor at a critical concentration of $x$ \cite{foot-critical-conc}.

For PMN itself an effective  Hamiltonian can  be expressed as
\begin{equation}
H^{B}_{PMN}=H^{B}_{PM^{*}_{1/3}N^{*}_{2/3}} + H^{extra},
\label{eq:H4}
\end{equation}
taking $x=1/3$.
Again a minimalist perspective is taken here for characterising $H^{extra}$ and investigating its principal consequences. We consider the effect of the different charges on Mg, Nb compared with the +4 charge on Mg* or Nb* and the corresponding modification of the interactions between the actual $B$-site ions. Coupling to the uniform strain is ignored here since we are concerned with the relaxor state which has no overall equilibrium polarization (unlike in ferroelectric PT). For simplicity, we shall assume that the $B$-sites are occupied independently randomly by Mg and Nb ions (in the ratio 1:2) \cite{foot-randomness}.

Before considering further the $B$-site perturbations we examine the expected effects of the alloying on the Pb and O ions. The leading perturbation on Pb ions due to the charge differences on the $B$-sites, compared with PM*N* (or  PT), is an attraction towards Mg neighbors (due  to their extra charge of of -2) and a repulsion from Nb neighbors (due to their extra charge +1), the former being twice as strong as the latter. In the na\"{i}ve (lowest approximation) paraelectric phase the $B$-neighbors are located at the equilibrium perovskite positions,  collinear pairwise along $<111>$ directions. The effects of pairs of collinear neighbours on an intermediate Pb cancel if both neighbors are of the same type, either both Mg or both Nb, leaving as the relevant perturbations those arising from a pair of dissimilar collinear neighbors  (one Mg and one Nb)  and of force-strength 3 
towards the Mg member (in units of $f_{\rm{Pb}}=e^{2}/{{\epsilon}R^{0}_{{\rm{Pb}}-B}}$ where $e$ is the electron charge, ${\epsilon}$ is the  electronic dielectric constant and $R_{{\rm{Pb}}-B}$ is the unperturbed Pb-$B$ separation). 
Thus the force distribution on Pb sites due to $B$-neighbors along any $<111>$ axis is 2/9 for each of $\pm  3$  units and 5/9 for no force at all. 
A simple vector sum over the contributions from the four $<111>$ axes at any Pb-site $\alpha$ gives the overall force ${\bf{f}_{\alpha}}$,
%with consequential  displacement shared between the Pb ion and the corresponding %neighboring $B$ ions.
yielding an equilibrium Pb displacement of $|{\bf{f}_{\alpha}}|/2\lambda_{\rm{Pb}}$ \cite{foot-B2}. 
Taking account only of nearest-neighbor Mg and Nb charges the $T=0$ distribution of Pb location displacements includes weight along all the $<111>$, $<100>$, $<110>$ and $<311>$ directions
 with magnitudes between $\sqrt{3}$ and 4
(in units of $3{ f_{\rm{Pb}}}/2\lambda^{\rm{Pb}}$)
%in appropriate units, 
as well as a small probability 
($\sim 0.1$) 
of no displacement. Further charges will spread the distribution further.
This could explain the observed quasi-spherical shell-like distribution of Pb displacements around the equilibrium perovskite positions at low temperatures \cite{Vakhrushev}, while at higher temperatures the standard Boltzmann weighting of energy shifts $\Delta{E}$, as proportional to $\exp{(\Delta{E}/k_{B}T)}$, leads to merging into a more Gaussian-like single peak.
% and a potential explanation of the observed Burns temperature \cite{Burns}. 
%a displacement of 3  ``length units'' 
%\footnote{
%of 
%$f_{\rm{Pb}} /2
%\lambda_{\rm{Pb}}$
%.} 
%, the ``units'' being temperature-dependent, determined by standard uncorrelated Boltzmann %statistics with the energy given by a \footnote{Of course in experiments there will also be the %usual Debye-Waller and experimental resolution broadening.}. 

Note that this appearance of a distribution shell of Pb displacement has nothing to do with the relaxor behaviour and, as presented above, is essentially simply a local effect of the statistical occupation of $B$-sites by Mg and Nb ions. In the lower temperature relaxor region the Nb and, to a lesser degree, Mg ions will be further displaced from their equilbrium perovskite locations and hence further modify the Pb distribution, but the modification is likely to be small by comparison with the main effect discussed above. 

An analagous star/shell-like  displacement distribution is, in principle,  expected for the O ions, which have 6 neighbours each along $<100>$ directions. In this case the effect  of Mg neighbours is repulsion and of Nb neighbours is attraction, relative to the O positions in the unperturbed perovskite, with effective displacement forces of 3, -3 and 0 O-units \cite{foot-O} of 2/9, 2/9 and 5/9 along the $<100>$ directions, again already without any cooperative ordering. Again this is dominantly a locally driven consequence of the random distribution of Mg and Nb, rather than related to relaxor behaviour.

Turning to the Mg and Nb sites,  
% RETURN TO THIS ISSUE
%the direct first order effects of the unperturbed surrounding 
%Pb and O 
%charges will cancel, although higher order perturbations will contribute, giving extra effective %(indirect)  $B$-site-to-site interactions. 
%There 
there will 
% will also 
be extra direct Coulombic $B$ site-to-site perturbations arising from the extra charges compared with the `bare' +4 of Mg*, Nb* or Ti; 
\begin{equation}
%H_{pert}^{B,B'}={\sum_{(ij)} e_{i}e_{j}/{{\epsilon}|({\bf{R}}^{0}_{i}+\bf{u}}_{i} -{\bf{R}}^{0}_{j} %-{\bf{u}}_{j})|}
H_{pert}^{B,B'}={\sum_{(ij)} e_{i}e_{j}/{{\epsilon}|({\bf{R}}^{0}_{ij}+\bf{u}}_{ij} |}
\label{eq:H5}
\end{equation}
where  ${\bf{R}}^{0}_{ij}=({\bf{R}}^{0}_{j} -{\bf{R}}^{0}_{i})$, ${\bf{u}}_{ij}=({\bf{u}}_{j} -{\bf{u}}_{i})$, ${\bf{R}}^{0}_{i}$ is the position of ion $i$ in the unperturbed system, $e_{i} =-2, +1$ respectively for sites $i$ occupied by Mg, Nb, and $\epsilon$ is the electronic dielectric constant.
Independently of whether the individual occupants of a pair of sites are Mg or Nb, the Coulomb energy favors an anti-ferroelectric interaction; both ions wishing to move closer together when they are of the same type, both wishing to move apart when they have opposite  signs of extra charge.

Considering only nearest neighbors explicitly for simplicity and expanding to first order in the $\{{u}\}$, the lowest-order extra displacement-influencing terms have the form
\begin{equation}
%H_{pert}^{B,B'(1)}={\sum_{(ij)} (e_{i}e_{j}}/{\epsilon}|{\bf{R}}^{0}_{i}-{\bf{R}}^{0}_{j}|^{3})
%{({\bf{u}}_{i} -{\bf{u}}_{j}).( {\bf{R}}^{0}_{i} -{\bf{R}}^{0}_{j})}.
H_{pert}^{B,B'(1)}={\sum_{(ij)} (e_{i}e_{j}}/{\epsilon}|{\bf{R}}^{0}_{ij}|^{3})
({{\bf{u}}_{ij}.{\bf{R}}^{0}_{ij})}.
\label{eq:H6}
\end{equation}
These are random field terms, albeit correlated with the specific $B$-site occupations by Mg and Nb. Since they are linear in the $\{\bf{u}\}$ they will necessarily lead to some displacement of the $B$-ions from their locations in pure perovskite. However, given that the harmonic restoring strength $\kappa$ of Mg is expected to be larger than that for Nb, the displacements will be more significant for the Nb than for the Mg.  

Higher order terms in eqn.($\ref{eq:H5}$) give further effective `exchange' interactions \cite{foot-sg-nom}, again quasi-random but correlated with the actual site-occupations, along with perturbations of effective $\kappa$ \cite{foot-second-order}.
%
%Further indirect interactions between $B$-sites will arise through the perturbations of shared %perturbed  Pb and O neighbours, again of effectively random-bond character. 

Thus the total effective $B$-$B$ Hamiltonian will contain extra `random-bond' interaction terms in addition to the spatially frustrated ones of  eqn.($\ref{eq:H3}$), albeit that still it is expected  that only the displacements of Nb sites will be relevant in the bootstrapping. Since the detailed character of the disorder and frustration is not important for the concept of spin glasses \cite{foot-sg}, if one ignores the random field terms, the combination of  exchange terms is expected to remain sufficient for the conceptual identification of relaxor behaviour as the analog of induced-moment soft pseudospin glass to continue, including the expectation of frequency-dependent permittivity peaks \cite{Tholence}.

The consequence of the random field terms for a true thermodynamic pseudo-spin glass transition is less clear. It is not possible to generate local random magnetic fields in experimental magnetic systems, so there are no direct experimental comparitors.  Theoretically, it is known that a sharp transition does still persist for the soluble infinite-ranged spin glass model of Sherrington and Kirkpatrick \cite{SK, AT, Sharma-Young}. Finite-ranged spin glass models are not exactly soluble. However, many workers believe that a field destroys a sharp transition  for a shorter-range Ising  model such as that of Edwards and Anderson \cite{EA} \cite{foot_GT}.  Furthermore, the situation for relaxors is probably intermediate between these two extremes because of the of long-range  Coulomb potentials and dipolar interactions. Studies of one-dimensional spin glasses with power-law decaying interaction strength, believed to emulate tunably both long- and short-range systems in different dimensions,  have demonstrated that for Ising spin-glass systems in a field \cite{Katzgraber} and also vector spin-glass  systems with quenched random fields \cite{Sharma-Young2} there exists a range of slow power-law decays over which an Almeida-Thouless \cite{AT} transition persists beyond just the infinite-ranged SK limit, while for more rapidly decaying interactions this is no longer the case. {Thus it is possible that the random fields in relaxors do not destroy the phase transition even in the complete thermodynamic sense, but a definitive conclusion is not currently available.} We note also that in computer simulations on a model for BZT analogous to eqn ($\ref{eq:H2}$)  but allowing also for tunable random fields and coupling to strain \cite{Akbarzadeh2012} it was found that
relaxor behavior was relatively unaffected by the random fields and strain.
However, it is anyway known from experiments on spin glasses that, in practice, permittivity peaks are still observed in small uniform applied fields  over finite experimental periods or finite frequency, so there could be finite-time relevance even without a tue thermodynamic equilibrium phase transition.

For further potential experimental clues as to the relative importance of the random field and random bond terms in leading to the relaxor behavior one can usefully consider  alloys $(PMN)_{1-y}(PT)_y$  in which there are 3 types of $B$-ion, Mg, Nb and Ti, and which we shall assume to be randomly distributed. By the same arguments as above the Mg are considered too immobile for the interactions to bootstrap spontaneous displacements but both Nb and Ti need to be included as having variable $\{{\bf{u}}_{i}\}$, albeit with different $\kappa$ and $\lambda$. Thus the fraction of effectively locally bootstrap-polarizable $B$-ions increases with $y$. 
%On the other hand, the strength of the effective random fields decreases with $y$.
Ignoring the differences between the $\kappa$-values of Ti and Nb, variations in the strength of the effective interaction term $H^{eff}_{int}$ with  $y$ and also the random field terms, leads to an expectation that the low $y$ behavior of $(PMN)_{1-y}(PT)_y$ will be relaxor with the onset temperature  increasing with $y$ \cite{foot-corollary}. This is precisely what is observed experimentally \cite{Colla} in studies of the frequency-dependent permittivity. As $y$ is increased the expectation is that the relaxor state will be supplanted  by ferroelectricity 
%(as in pure PT)
 at a critical $y_c$ \cite{foot-strain}, with $T_f$ continuing to increase with $y$. Again this is as observed \cite{Colla}.

Let us now turn to the random fields. Since the density of excess charges decreases with $y$ on alloying to $(PMN)_{1-y}(PT)_y$ the strength of random fields decreases with $y$. If the origin of relaxor behavior is believed to lie in random fields, one might thus expect that on dilution of PMN with PT  the relaxor onset temperature would  decrease with $y$. However, this is in contrast to experimental observations \cite{Colla}. We are therefore led to suspect that relaxor behavior is dominantly of effective random-bond origin \cite{foot-random-bond}, with random fields playing only a secondary role.
%{\textcolor{red}{Reality may however lie between the random-bond and random-field pictures.}}

In \cite{DS13} an extension of the modeling of eqn.(\ref{eq:H2}) was used to explain the origin and character of nanodomains through a mapping to a related Anderson localization model. The general concepts of that analogy apply here too. 
%, albeit that terms linear in $u$, such as those in eqn.(\ref{eq:H6}), induce a  random-field %induced `random ripple' displacement background,  beyond the simple Anderson eigenvalue %equation mapping. 
 More specifically, $H^{B}_{PMN}$   is re-interpreted as a (microscopic and heterogeneous) Ginzburg-Landau-like free energy with temperature-renormalized weighting 
of the coefficients and the interaction, denoted below by tildes. 
%with the most important temperature-dependent  term for the present discussion being that of 
%, especially for 
% the $\kappa$ \cite{foot-GL}.  
Mean-field solutions are given by minimization with respect to the $\{\bf{u}\}$;
\begin{equation}
{\tilde\kappa_{i}} {\bf{u_{i}}} 
+2{\tilde\lambda_{i}} {\bf{u_{i}}} {|\bf{u_{i}}|}^2
+ {\sum_{j} \partial { \tilde H^{B}({\bf{u}}_{i}, {\bf{u}}_{j},  f({\bf{R}}_{ij})}/\partial{\bf{u}}_{i} }=0.
\label{self-cons}
\end{equation}
%
%where the tildes are to denote the $T$-renomalization. 
The most important  temperature-dependent  term for the present discussion is that of the \{$\tilde\kappa$\} which  increase with temperature, and control the phase transition, just as  does the coefficient of the term quadratic in the  order parameter  in a conventional Ginzburg-Landau free energy.

The  terms in $\tilde H^{B}$ that are linear but random  in $\{{\bf{u}}\}$  give rise to a backgound `random ripple'. But of more interest are the interaction terms that enable internally ordered clusters identifiable as `nanodomains'. As pointed out in \cite{DS13} these can be conceptually `identified' with eigenfunctions of an analogous Anderson equation, with the nanodomains  corresponding to negative energy localized states of the eigenequation and the phase transition  itself the analog of the energy of the mobility edge crossing a critical value and yielding a macroscopic extended state. 

The conceptual procedure has been described in \cite{DS13} but is re-iterated for possible convenience. Consider for illustrative purposes a simple scalar analog of eqn.($\ref{self-cons}$) 
\begin{equation}
\tilde\kappa_{i} u_{i} 
+2 \tilde\lambda_{i} u_{i}^3
- \sum_{j} \tilde J_{ij} u_{j}  -\sum_{i} \tilde h_{i}=0.
\label{simplified_sc}
\end{equation}
and compare it with an Anderson-type eigen-equation
\begin{equation}
\tilde\kappa_{i} \phi_{i} 
-\sum_{j} \tilde J_{ij} \phi_{j} =E\phi_{i}.
\label{Anderson_equiv}
\end{equation}
If the random-field \{$\tilde h_{i}$\} are ignored then there is a parallel between these equations that identifies approximate solutions of eqn. ($\ref{simplified_sc}$) in terms of the  eigenfunctions and eigenvalues of eqn ($\ref{Anderson_equiv}$), given by  
\begin{eqnarray}
%\left{
 m_{i}=&0,\\    \nonumber
&\delta_{i}{\sqrt{-E/2\tilde \lambda_{i}}~~~;~ E\le 0} 
%\right}.
\end{eqnarray}
where 
\begin{eqnarray}
\delta_{i}=&1;~~\phi_{i} \neq 0 \\ \nonumber
&0;~~\phi_{i}=0.
\end{eqnarray}
Nanodomains are then given by negative energy solutions of the Anderson equation. Cooperative order  also requires that there be extended state solutions but for disordered \{$\tilde{\kappa}$\} the Anderson equation has localised states at its extremities (and possibly everywhere).  Thus those nanodomains are truly frozen cooperatively only if the Anderson equation has extended solutions and that its lower mobility edge is negative. Since the density of states moves linearly with shifts of the $\tilde \kappa$ distribution and that distribution moves continuously to higher values as the temperature increases, it follows that, within the realm of ignoring the random field terms,  the onset of metastable nanodomains as the temperature is reduced can be associated with the passage of the density of states across zero and the phase transition to a cooperative phase with the passage of the lower mobility edge across zero. Whether the phase transition is to a ferroelectric or relaxor phase depends on the disorder, as discussed in \cite{DS13} and briefly above \cite{transition-type}. 

Of course these details have to be modified when account is taken of the random field terms in 
eqn. ($\ref{simplified_sc}$); for example, the fields will always induce some displacement and the mapping is no longer precise. However, the qualitative concept of localized metastable nanodomains growing and coalescing remains.

%As commented upon in \cite{DS13} the more conventional `picture' of nanodomains coalescing %at the relaxor-onset temperature can be mimicked qualitatively through a mapping to  %mesovariables corresponding to localized quasi-eigenstates, with effective interactions between %them leading to macroscopic correlation as However, in the picture presented above the %relevant mesovariables mainly involve the Nb ions and not the Pb, O or Mg ions.

The pseudo-spin glass model employed above invoked the site randomness of relevant ions together with a longish-ranged interaction frustrated as a function of separation and made comparison with what is known about more conventional experimental spin glasses. However, once one accepts the premise that it is the combination of frustration and quenched disorder that underlies the existence of a non-periodic phase with interesting non-ergodicity, then one can pass to potentially more computationally convenient theoretical encapsulation in models with random-bond disorder, as epitomised in magnetic spin glasses by the Edwards-Anderson model \cite{EA}, extended to soft spins as above and with a ferromagnetic bias of the bond distribution \cite{SS}\cite{not-SK}. 

In conclusion, we have argued that PMN, like BZT, has as its underlying core for relaxor behavior an effective induced-moment soft pseudo-spin glass on  $B$-sites of the $\mathrm{ABO}_{3}$ perovskite structure, in particular of the Nb ions, albeit with distracting effects due to the charge differences of the Mg and Nb ions, such as displacements on the Pb sites. This analogy suggests that PMN (and some other relaxors) should exhibit several further spin-glass-like properties. However, it should be noted  that to demonstrate this assertion we have employed several modeling simplifications and also that  it is known that relaxor materials, including PMN, exhibit a number of other transitions and features. Thus there is no claim here  for a complete explanation, but rather an argument for  an underlying conceptual and pictorial basis that is significantly different from the conventional starting perspectives for these materials. {Complete reality, however, may involve elements of both the B-dominated driving discussed above and more conventional aspect of driving via the Pb ions.}  The hope is that this {unconventional perspective} will  stimulate other fruitful  investigations, extensions and applications, both theoretical and experimental \cite{suggestion}. Clearly, we might expect similar behaviors for other displacive ionic alloys, but they are not explored further here. 

We might recall that we have assumed random locations of substituted $B$-ions. If there is chemical periodic ordering then there is less randomness and more opportunity for a periodic optimal compromise of cooperative ordering. This was observed long ago \cite{Stenger} and also confirmed in a more recent computational study \cite{Hemphill}. 

%A similar demonstration of the necessity to move from simple crystalline structure to yield %pseudo-spin-glass behaviour was also demonstrated recently in disordered martensitic %systems, such as ${\bf{Ti}}_{1-x}{\bf{Ni}}_{1+x}$  \cite{Sherrington_martensite} \cite{Ren}, 

%A  complementary approach to ${\rm{PMN}}_{0.75}{\rm{PT}}_{0.25}$ using first-principles and %molecular dynamics simulations and questioning the conventional PNR picture has recently %been considered in\cite{Takenaka}.

\section*{Acknowledgements}

The author thanks Prof. Roger Cowley for stimulating his interest in relaxors and Prof. Rasa Pirc for drawing his attention to the paper of Akbarzadeh et al on BZT. He also thanks Profs. Wolfgang Kleemann and Laurent Bellaiche for comments and references in connection with \cite{DS13}. He is grateful to the Leverhulme Trust for the award of an Emeritus Fellowship.


\begin{thebibliography}{99}
\bibitem{foot-isovalent}{Here we are considering the valence of the $B$-site ions.}
\bibitem{foot-B} {Note that the expressions ‘$B$-site’ or ‘$B$-type ion’ are used
to refer to the $B$ in $AB\mathrm{O}_{3}$ and not the abbreviation B for
Ba in BZT.}
\bibitem{foot-sg}{Note that the crucial ingredients for potential spin glass behaviour are frustrated interactions and quenched disorder, but also 
that within that `envelope' there exist many different specific instances. }
\bibitem{foot-critical-conc}{More particularly there are expected to be  two critical concentrations, $x_{c1}$ and $x_{c2}<x_{c1}$, with ferroelectricity for $x>x_{c1}cc$, relaxor behavior for $x_{c1}>x_{c2}$ and only paraelectricity for $x<x_{c2}$.}
\bibitem{foot-randomness}{This is in natural accord with the philosophy of \cite{DS13};  it has also  been observed empirically that randomness is advantageous for relaxor formation, whereas periodicity favours more conventional ordered states.}
\bibitem{foot-B2} {Note that the $B', B''$ ions sit in symmetric environments, so that the extra forces on them from the surrounding $\rm{Pb}^{++}$ and $\rm{O}^{--}$, due to their excess charges, will cancel to lowest order.}
\bibitem{foot-O}{For O-units $\rm{R_{Pb-B}}$ is repaced by $\rm{R_{Pb-O}}$ and in assessing the positional shifts the O spring constant is employed.}
\bibitem{foot-corollary} {A corollary is that alloying PMN with ${\mathrm {PbZrO}}_3$ will reduce the relaxor transition temperature and the size of the permittivity peak.}
\bibitem{foot-strain}{As is well-known, coupling to the strain leads to a significant  contribution to the energy minimization in the ferroelectric phase of PT, with resultant accompaniment of the ferroelectric transition by the  onset of global spontaneous strain.  A similar spontaneous strain is predicted in the ferroelectric regime of $(PMN)_{1-x}(PT)_x$. However, it is ineffective in the relaxor phase because of the quasi-randomness of the local B-displacements.}
\bibitem{foot-random-bond}{Here we are including the combination of site-to-site $u$-frustration as a function of separation and random-site occupation within the nomenclature `effective random-bond', in distinction from `random-field' terms linear (or of odd local order) in $u$.}
\bibitem{foot-GL}{The renormalized $\tilde\kappa$ will be increasing functions of temperature, just as for the coefficients of quadratic order parameter terms in conventional Ginzburg-Landau.}
\bibitem{foot_GT} 
{For vector spins in a uniform field there is expected still to be a sharp transition to transverse spin glass order \cite{GT}}
\bibitem{transition-type}
{For PMN the transition is to relaxor but for PMN-PT it becomes feroelectric beyond a critical PT concentration.}
\end{thebibliography}
\end{document}